\documentclass[aps,prl,twocolumn,superscriptaddress]{revtex4-2}
\usepackage{babel}
\usepackage{bbm}
\usepackage{mathrsfs}
\usepackage{amsmath}
\usepackage{amsfonts}
\usepackage[colorlinks=true,linkcolor=blue,urlcolor=blue,citecolor=blue,anchorcolor=blue]{hyperref}
\usepackage{graphicx}
\usepackage{subfigure}
\usepackage{epsfig}
\usepackage{dcolumn}
\usepackage{bm}
\usepackage{color}
\usepackage{natbib}
\usepackage{amssymb}
\usepackage{xcolor}
\usepackage{braket}

\usepackage{float}
\usepackage{lipsum}
\usepackage{pifont}
\usepackage[normalem]{ulem}

\definecolor{newtxtcolor1}{rgb}{0.8, 0, 0.2}
 
\definecolor{ngreen}{rgb}{0.2,0.7,0.2}
\definecolor{nred}{rgb}{0.9,0.1,0}

\definecolor{npurple}{rgb}{0.6, 0.3, 0.8} 
\newcommand{\nx}[1]{{\textcolor{npurple}{#1}}}

\begin{document}

\title{Noncommutativity as a Universal Characterization for Enhanced Quantum Metrology}

\author{Ningxin Kong}
\affiliation{\mbox{State Key Laboratory for Mesoscopic Physics, School of Physics, Frontiers Science Center for Nano-optoelectronics,} $\&$ Collaborative Innovation Center of Quantum Matter, Peking University, Beijing 100871, China}

\author{Haojie Wang}
\affiliation{\mbox{State Key Laboratory for Mesoscopic Physics, School of Physics, Frontiers Science Center for Nano-optoelectronics,} $\&$ Collaborative Innovation Center of Quantum Matter, Peking University, Beijing 100871, China}
 \affiliation{Hefei National Laboratory, Hefei 230088, China}

\author{Mingsheng Tian}
\thanks{Current address: Department of Physics, The Pennsylvania State University, University Park, Pennsylvania, 16802, USA}
\affiliation{\mbox{State Key Laboratory for Mesoscopic Physics, School of Physics, Frontiers Science Center for Nano-optoelectronics,} $\&$ Collaborative Innovation Center of Quantum Matter, Peking University, Beijing 100871, China}

\author{Yilun Xu}
\affiliation{\mbox{State Key Laboratory for Mesoscopic Physics, School of Physics, Frontiers Science Center for Nano-optoelectronics,} $\&$ Collaborative Innovation Center of Quantum Matter, Peking University, Beijing 100871, China}

\author{Geng Chen}
\affiliation{Laboratory of Quantum Information, University of Science and Technology of China, Hefei 230026, China}

\author{Yu Xiang}
\email{xiangy@xjtu.edu.cn}
\affiliation{Ministry of Education Key Laboratory for Nonequilibrium Synthesis and Modulation of Condensed Matter, Shaanxi Province Key Laboratory of Quantum Information and Quantum Optoelectronic Devices, School of Physics, Xi'an Jiaotong University, Xi'an 710049, China}

\author{Qiongyi He}
\email{qiongyihe@pku.edu.cn}
\affiliation{\mbox{State Key Laboratory for Mesoscopic Physics, School of Physics, Frontiers Science Center for Nano-optoelectronics,} $\&$ Collaborative Innovation Center of Quantum Matter, Peking University, Beijing 100871, China}
 \affiliation{Hefei National Laboratory, Hefei 230088, China}
\affiliation{\mbox{Collaborative Innovation Center of Extreme Optics, Shanxi University, Taiyuan, Shanxi 030006, China}}

\begin{abstract} 
A central challenge in quantum metrology is to effectively harness quantum resources to surpass classical precision bounds. Although recent studies suggest that the indefinite causal order may enable sensitivities to attain the super-Heisenberg scaling, the physical origins of such enhancements remain elusive.
Here, we introduce the nilpotency index $\mathcal{K}$, which quantifies the depth of noncommutativity between operators during the encoding process, can act as a fundamental parameter governing quantum-enhanced sensing. 
We show that a finite $\mathcal{K}$ yields an enhanced scaling of root-mean-square error as $N^{-(1+\mathcal{K})}$. 
Meanwhile, the requirement for indefinite causal order arises only when the nested commutators become constant.
Remarkably, in the limit $\mathcal{K} \to \infty$, exponential precision scaling $N^{-1}e^{-N}$ is achievable. 
We propose experimentally feasible protocols implementing these mechanisms, providing a systematic pathway towards practical quantum-enhanced metrology.
\end{abstract}
    
    \maketitle

\begin{figure}
    \centering
    \includegraphics[width=1\linewidth]{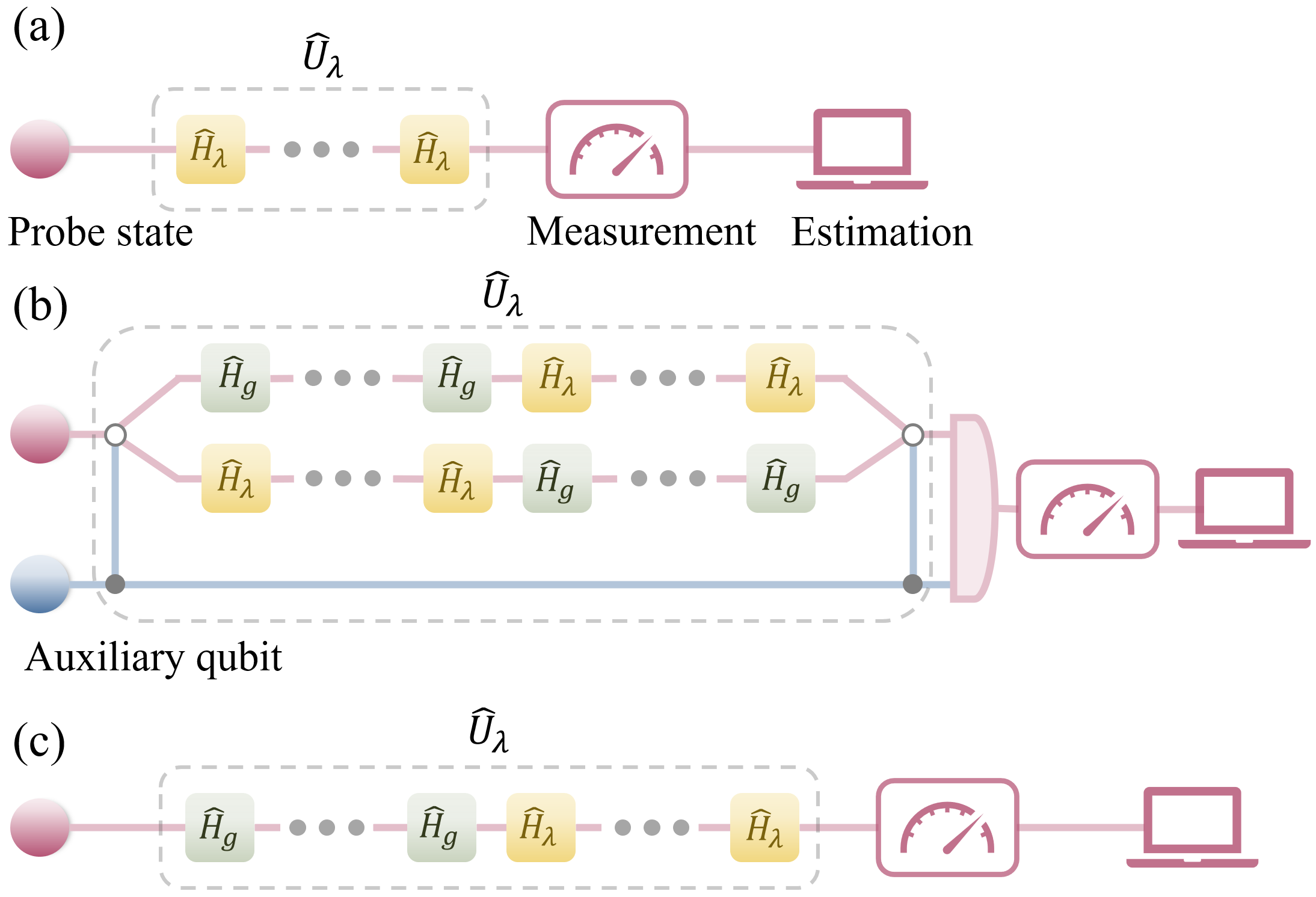}
    \caption{The schematic illustrates representative quantum estimation schemes: (a) In the sequential protocol, a single probe state sequentially undergoes $N$ times of the parameter-encoding operator $\hat{H}_\lambda$. (b) In the ICO-based scheme, the probe evolves through a noncommutative encoding sequence whose order is coherently controlled by an auxiliary qubit. (c) In the general noncommutative encoding framework, the probe undergoes a sequence of $2N$  noncommutative encoding operations—$N$ times $\hat{H}_g$ and $N$ times $\hat{H}_\lambda$—executed in a definite order.}
    \label{fig1}
\end{figure}

\textit{Introduction.---}Quantum metrology represents one of the most promising near-term applications of quantum technologies~\cite{Maccone2004,Maccone2006,Romalis2007,PARIS2009,Cappellaro2017,Treutlein2018}. Its fundamental advantage stems from the ability of quantum resources—such as entanglement and coherence—to enhance measurement precision beyond the limits achievable with classical methods~\cite{Maccone2011,Zoller2024,Zelevinsky2024}. 
This is exemplified by the reduction in the root-mean-square error (RMSE) for parameter estimation from the standard quantum limit, scaling as $N^{-1/2}$, to the Heisenberg limit, scaling as $N^{-1}$, when utilizing $N$ resources~\cite{Geremia2007,Berry2020}.
Such Heisenberg scaling has been demonstrated across diverse experimental platforms, achieved either by preparing $N$ entangled probes measured in parallel~\cite{Zeilinger2004,Silberberg2010,Pan2023} or by employing a single probe that evolves through and is measured after $N$ repetitions [Fig.~\ref{fig1}(a)]~\cite{Pryde2007,Guo2018,Guo2021}.
 
To further push the sensitivity limit in measurement precision, recent processes involving indefinite causal order (ICO), where the sequence of fundamental encoding operations lacks definite temporal sequences, have been incorporated into quantum metrology~\cite{Walther2024,Brukner2014,Brukner2012,Rafa2023,Yang2023,Alastair2024}. 
A paradigmatic example is the quantum SWITCH [Fig.~\ref{fig1}(b)], in which a control quantum system coherently dictates the order of two operations applied to a target system~\cite{Valiron2013,Walther2015,Walther2017,White2018,Giulio2024,Brukner2016,Brukner2016njp}. 
Significantly, using state-of-the-art photonic techniques, a precision scaling with super-Heisenberg limit $N^{-2}$ has been demonstrated~\cite{Liu2025,Guo2023}.

However, the fundamental mechanism by which ICO enhances measurement precision remains poorly understood, representing a key foundational question~\cite{Giulio2020,Guo2023,Zeng2024,Frey2019,Li2024,Ban2023,Delgado2023}. Understanding the essential resource underpinning this advantage holds significant promise for reducing experimental overhead while advancing the ultimate sensitivity limits.
While it is often assumed that the coherence of causal order plays a central role, evidence suggests that simply increasing the number of orders does not necessarily lead to improved precision~\cite{footnote}. 
In existing protocols performing displacements generated by a given quadrature $\hat{X}$ and the conjugate quadrature $\hat{P}$~\cite{Giulio2020,Guo2023,Liu2025} or two unitary gates $\hat{U}$ and $\hat{U}^\dagger$ for parameter encoding~\cite{Zeng2024}, continuing to add similar paths brings little benefit and even makes the experiment harder to implement.
Instead, the enhancement appears to stem from the noncommutativity between encoding operations in two paths, highlighting it as a more critical factor than only order multiplicity.

Here, we establish a quantum metrology framework that exploits noncommutative sequences of encoding operations to enhance measurement precision. Central to our approach is the introduction of the noncommutativity depth $\mathcal{K}$, which is defined as the maximal integer yielding a nonzero nested commutator and serves as a fundamental quantity characterizing the noncommutative structure of the encoding operators. We rigorously demonstrate that $\mathcal{K}$ universally governs the scaling behavior of RMSE.
For finite $\mathcal{K}$, the RMSE scales as $N^{-(1+\mathcal{K})}$ when the nested commutator remains nonconstant, enabling precision enhancement even without requiring the ICO scheme. However, when the $\mathcal{K}$-th order commutator becomes constant, ICO becomes necessary to unlock the optimal scaling. In the asymptotic limit $\mathcal{K}\to\infty$, we show that an exponential RMSE scaling $N^{-1}e^{-N}$ can be achieved, provided the nested commutators exhibit nonzero variance on the relevant initial state.
Furthermore, we propose experimentally feasible protocols capable of achieving these enhanced scalings. Our findings offer a systematic approach to designing quantum control strategies for metrological optimization under realistic experimental conditions.

\textit{General framework.---}Generally, the precision of quantum metrology estimation governed by a unitary encoding $\hat{U}_\lambda = \text{exp}(-i\lambda \hat{H}_\lambda)$ is fundamentally determined by the distinguishability between the neighboring output states $|\Psi_\lambda\rangle = \hat{U}_\lambda|\Psi\rangle$ and $|\Psi_{\lambda+\delta\lambda}\rangle = \hat{U}_{\lambda + \delta\lambda}|\Psi\rangle$. This distinguishability is quantified by the quantum Fisher information (QFI) $F_\lambda = 4\text{Var}[\hat{h}_\lambda]_{|\Psi\rangle}$, where $\hat{h}_\lambda = i\hat{U}_\lambda^\dagger \partial_\lambda \hat{U}_\lambda$ denotes the local generator associated with $\lambda$, and the variance $\text{Var}[\cdots]_{|\Psi\rangle}$ is evaluated to the probe state $|\Psi\rangle$. The QFI sets a fundamental lower bound on the RMSE via the quantum Cramér-Rao bound (QCRB) $\Delta\lambda = 1/\sqrt{\nu F_\lambda}$, where $\nu$ is the number of independent measurements. This bound is saturable in the asymptotic limit $\nu \to \infty$ through optimal choice of measurement and estimator~\cite{Cai2021,Caves1994,Augusto2009,Kok2010}. 

We begin by introducing the general framework of the noncommutative metrology scheme [Fig.~\ref{fig1}(c)]. The encoding unitary $\hat{U}_\lambda$ is constructed by sequentially applying two noncommutative operators, $\hat{H}_\lambda$ and $\hat{H}_g$ (i.e., $[\hat{H}_g,\hat{H}_\lambda] \neq 0$), and takes the form
\begin{equation}
        \hat{U}_\lambda = \prod_{j=1}^{N_\lambda} e^{-i\lambda_j\hat{H}_\lambda} \prod_{i=1}^{N_g} e^{-ig_i\hat{H}_g} = e^{-iN_\lambda\overline{\lambda} \hat{H}_\lambda} e^{-iN_g \overline{g} \hat{H}_g},
        \label{eq1}
\end{equation}
where $\overline{\lambda} = \sum_{j=1}^{N_\lambda} \lambda_j/N_\lambda$ is the target parameter to be estimated, and $\overline{g} = \sum_{i=1}^{N_g} g_i/N_g$ is an auxiliary control parameter. The probe experiences $N_\lambda$ and $N_g$ applications of $\hat{H}_\lambda$ and $\hat{H}_g$, respectively, following a definite operator ordering.

To quantify the depth of noncommutativity
between $\hat{H}_\lambda$ and $\hat{H}_g$, we define the \textit{nilpotency index} $\mathcal{K}\in\mathbb{N}$ via the adjoint action $\mathrm{ad}_{\hat{A}}(\hat{B}) \equiv [\hat{A},\hat{B}]$, such that
\begin{equation}
    (\mathrm{ad}_{\hat{H}_g})^n (\hat{H}_\lambda) \equiv \underbrace{[\hat H_g,\ldots[\hat H_g}_{n\;\text{times}},\hat H_\lambda]\ldots].
\end{equation}
We then define $\mathcal{K}$ as the maximal integer for which the nested commutator remains nonzero:
\begin{equation}
    \mathcal{K} \equiv \max \left\{ n \in \mathbb{N} \;\middle|\; (\mathrm{ad}_{\hat{H}_g})^{n}(\hat{H}_\lambda) \neq 0 \right\}.
\end{equation}
That is, the action of $\mathrm{ad}_{\hat{H}_g}$ on $\hat{H}_\lambda$ terminates at order $\mathcal{K}+1$, $(\mathrm{ad}_{\hat{H}_g})^{\mathcal{K}+1}(\hat{H}_\lambda) = 0,$ which implies that $\mathrm{ad}_{\hat{H}_g}$ is nilpotent of index $\mathcal{K} + 1$ on $\hat{H}_\lambda$. We note that the nilpotency index $\mathcal{K}$ is determined by the noncommutative algebraic structure and is independent of how many times the encoding operators are applied sequentially.

The QFI for estimating $\overline{\lambda}$ from the evolved state $|\Psi_\lambda\rangle = \hat{U}_\lambda |\Psi\rangle$ is determined by the local generator:
\begin{equation}
    \hat{h}_{\overline{\lambda}} = i\hat{U}_{\lambda}^\dagger \partial_{\overline{\lambda}} \hat{U}_{\lambda} = N_\lambda \sum_{n=0}^{\mathcal{K}} \frac{(iN_g\overline{g})^n}{n!} [\hat{H}_g,\hat{H}_\lambda]_{(n)},
    \label{eq2}
\end{equation} 
where $[\hat{H}_g,\hat{H}_\lambda]_{(n)}$ denotes the $n\text{-th}$ order nested commutator with $[\hat{H}_g,\hat{H}_\lambda]_{(0)}=\hat{H}_\lambda$. Assuming $N_\lambda = N_g = N$ without loss of generality, Eq.~(\ref{eq2}) shows that the local generator includes terms scaling as $N^{1+\mathcal{K}}$, with the nilpotency index $\mathcal{K}$ directly determining the scaling behavior of the QFI.
This observation naturally motivates two distinct metrological regimes: one where $\mathcal{K}$ is finite, and another where it diverges to infinity.

\textit{Finite \textit{nilpotency index} $\mathcal{K}$.---}For the case of finite nilpotency index 
$\mathcal{K}$, the QFI of system is given by $F_{\bar{\lambda}} = 4 N^2 
\sum_{m,n=0}^{\mathcal{K}} \frac{(iN \overline{g})^{m+n}}{m!n!} (\langle [\hat{H}_g,\hat{H}_\lambda]_{(m)} [\hat{H}_g,\hat{H}_\lambda]_{(n)} \rangle - \langle [\hat{H}_g,\hat{H}_\lambda]_{(m)} \rangle \langle [\hat{H}_g,\hat{H}_\lambda]_{(n)}\rangle)$.
In the asymptotic regime of large $N$, the leading-order scaling of QFI for the parameter $\overline{\lambda}$ takes the form
\begin{equation}
    F_{\overline{\lambda}} \simeq 4 \frac{N^{2(1+\mathcal{K})}}{(\mathcal{K}!)^2} \overline{g}^{2\mathcal{K}} \text{Var}[[\hat{H}_g,\hat{H}_\lambda]_{(\mathcal{K})}]_{|\Psi\rangle}.
\end{equation}
Under the conditions of $\overline{g} \neq 0$ and a nonzero variance $\text{Var}[[\hat{H}_g,\hat{H}_\lambda]_{(\mathcal{K})}]_{|\Psi\rangle}\neq 0$ in the initial state, the QFI scales as $F_{\overline{\lambda}} \simeq N^{2(1+\mathcal{K})} / (\mathcal{K}!)^2$, establishing a direct and quantitative connection between the nilpotency index $\mathcal{K}$ and the degree of metrological enhancement.
Through the QCRB, this scaling yields an RMSE $\Delta \overline{\lambda} \sim N^{-(1+\mathcal{K})}\mathcal{K}!$.
This enhancement arises solely from the intrinsic noncommutativity between the auxiliary operator $\hat{H}_g$ and the parameter-encoding operator $\hat{H}_\lambda$, and requires no exotic quantum resources such as superpositions of causal orders that are often invoked in ICO-based schemes.
Notably, the standard Heisenberg limit $\Delta \overline{\lambda} \sim N^{-1}$ emerges as a special case corresponding to $\mathcal{K} = 0$, where no noncommutative structure is present. 

\begin{figure}[t]
    \centering
    \includegraphics[width=1\linewidth]{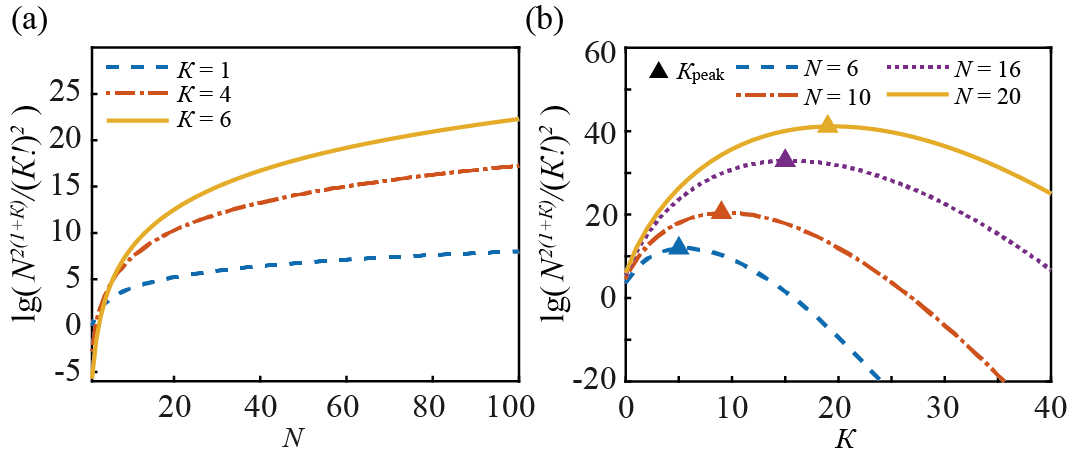}
    \caption{(a) Logarithmic scaling of the coefficient of the leading-order QFI term ${N^{2(1+\mathcal{K})}}/{(\mathcal{K}!)^2}$, as a function of the number of operations $N$, for fixed nilpotency indices $\mathcal{K} = 1$, 4, and 6. Each curve represents the leading-order contribution from the series expansion of the total QFI in its respective encoding scheme, and exhibits a scaling $F_{\overline{\lambda}}^{(\mathcal{K})} \sim N^{2(1+\mathcal{K})}$, consistent with the structure of nested commutators.
    (b) Logarithmic scaling of the leading-order QFI coefficient as a function of the nilpotency index $\mathcal{K}$, for fixed $N = 6$, 10, 16, and 20. For each given $N$,  %an optimal 
    a peak value $\mathcal{K}_{\mathrm{peak}}$ emerges  (see Supplemental Material for its detailed expression~\cite{sm}) which maximizes the contribution to the total QFI.} 
    \label{fig2}
\end{figure}

As illustrated in Fig.~\ref{fig2}(a), in the asymptotic regime of large $N$ ($N \gg \mathcal{K}$), the leading-order scaling of QFI increases with the number of operations $N$ and the nilpotency index $\mathcal{K}$. 
But in practical implementations, the number of operations $N$ is inevitably limited by technical constraints. This introduces a trade-off in the scaling behavior governed by the nilpotency index $\mathcal{K}$: increasing $\mathcal{K}$ enhances the sensitivity via the power-law term, while the factorial term $(\mathcal{K}!)^{-2}$ leads to a suppression.
As shown in Fig.~\ref{fig2}(b), this competition results in a peak nilpotency index $\mathcal{K}_{\text{peak}}$ for a given number of operations $N$. 
Thus, under limited experimental resources, $\mathcal{K}_{\text{peak}}$ provides a practical guideline for designing the encoding process to obtain significant metrological gain. Beyond this point, further increasing $\mathcal{K}$ yields only marginal improvements in the total QFI.

However, when the $\mathcal{K}$-th order nested commutator becomes constant, for instance, in the case $\hat{H}_\lambda = \hat{P}= i(\hat{a}^\dagger-\hat{a})/\sqrt{2}$ and $\hat{H}_g = \hat{X}=(\hat{a}^\dagger+\hat{a})/\sqrt{2}$, where $\mathcal{K}=1$, the nested commutator $[\hat{H}_g,\hat{H}_\lambda]=i$, and hence $\text{Var}[[\hat{H}_g,\hat{H}_\lambda]_{(\mathcal{K}=1)}]_{|\Psi\rangle} = 0$, the RMSE scaling reduces to $\Delta \overline{\lambda} \sim N^{-1}$, recovering the standard Heisenberg limit. In this scenario, ICO-based schemes can unlock the enhanced scaling $\Delta \overline{\lambda} \sim N^{-(1+\mathcal{K})}\mathcal{K}!$ by introducing coherent superpositions of causal orders, which act as additional quantum resources to restore the contribution of the $\mathcal{K}$-th order nested commutator. Importantly, even under ICO, the metrological advantage is bounded by the underlying noncommutative structure between $\hat{H}_g$ and $\hat{H}_\lambda$~\cite{sm}. While ICO protocols cannot surpass the $\mathcal{K}$-dependent scaling limit, they can activate metrological enhancement that is otherwise inaccessible under definite causal order by coherently exploiting quantum control over operator sequences.

More broadly, an ICO-based scheme can be used to estimate either $\overline{\lambda}$ or $\overline{g}$. However, when the associated nilpotency index differ, i.e., $\mathcal{K}_{[\hat{H}_g, \hat{H}_\lambda]} \neq \mathcal{K}_{[\hat{H}_\lambda, \hat{H}_g]}$, the achievable precision for each parameter becomes intrinsically different. In such cases, the RMSE for each estimate is dictated by the nilpotency index corresponding to the relevant generator, depending on the asymmetric noncommutative structure of the encoding process~\cite{sm}.

\textit{Infinite nilpotency index $\mathcal{K} \to \infty$.---}In this regime, by constructing a class of noncommutative encoding operators, we demonstrate that the infinite nilpotency can induce an exponential precision scaling of RMSE.

Specifically, we consider a family of Hamiltonians $\hat{H}_\lambda$ and $\hat{H}_g$ satisfying the commutation relation: $[\hat{H}_g,\hat{\Lambda}] = i\sqrt{p}\hat{\Lambda}$,
where $\hat{\Lambda} = i\sqrt{p}\hat{C}+\hat{D}$, with $\hat{C} = [\hat{H}_g,\hat{H}_\lambda]$ and $\hat{D} = [\hat{H}_g,\hat{C}]$. Here, $p >0$ is a parameter-independent constant determined
by $\hat{H}_\lambda$ and $\hat{H}_g$.
This algebraic structure gives rise to a closed-form expression for the local generator $\hat{h}_{\overline{\lambda}}$, obtained via a Taylor expansion from Eq.~(\ref{eq2}):
\begin{align}
    \hat{h}_{\overline{\lambda}} & = N\hat{H}_\lambda + N\frac{\sinh(N\overline{g}\sqrt{p})}{\sqrt{p}} i\hat{C} + N\frac{1-\cosh(N\overline{g}\sqrt{p})}{p}\hat{D},
    \label{eq6}
\end{align}
revealing an exponential dependence on $N$ when $\text{Re}(\overline{g}) \neq 0$.
As a result, the QFI is well quantified and can achieve an exponential enhancement:
\begin{equation}
    F_{\overline{\lambda}} \propto N^2 e^{2N\overline{g}\sqrt{p}} (\frac{\text{Var}[i\hat{C}]_{|\Psi\rangle}}{p} + \frac{\text{Var}[\hat{D}]_{|\Psi\rangle}}{p^2} + \frac{\text{Cov}[i\hat{C},\hat{D}]_{|\Psi\rangle}}{p^{3/2}}),
    \label{eq7}
\end{equation}
demonstrating $\Delta\overline{\lambda} \simeq N^{-1} e^{-N}$ scaling driven by the underlying noncommutative structure of the encoding operators. 

Importantly, this enhancement is generic and does not require specially prepared probe states: exponential scaling holds for a variety of $|\Psi\rangle$ as long as $\text{Var}[i\hat{C}]_{|\Psi\rangle}\neq 0$,  
$\text{Var}[\hat{D}]_{|\Psi\rangle}\neq 0$, 
or $\text{Cov}[i\hat{C},\hat{D}]_{|\Psi\rangle} \neq 0$.
This framework highlights the role of infinite nilpotency index $\mathcal{K} \to \infty$ as a necessary condition for exponential enhancement. Furthermore, it reveals the potential for systematically optimizing metrological performance by engineering the auxiliary operator $\hat{H}_g$ to tailor the noncommutative structure, for both finite and infinite $\mathcal{K}$.

\textit{Examples}.---We now propose several experimentally feasible protocols, implemented within continuous-variable (CV) systems, to realize the predicted precision enhancement. 
This focus stems from a fundamental limitation in discrete-variable (DV) systems: as rigorously proven in the Supplementary Material~\cite{sm}, the variance of the local generator is strictly bounded by the finite geometry of the generalized Bloch vector space for a $d$-dimensional system. This bound inherently precludes both the power-law ($N^{-(1+\mathcal{K})}$) and exponential ($N^{-1}e^{-N}$) scalings achievable in CV systems, limiting precision to at most oscillatory behavior with $N$. Consequently, ICO cannot be used to achieve super-Heisenberg scaling for DV quantum metrology, addressing an interesting open question in Ref.~\cite{Guo2023}.

(i) Finite nilpotency index $\mathcal{K}$.---Now we present a quantum metrological protocol that achieves enhanced scaling by exploiting noncommutative operations with a finite nilpotency index $\mathcal{K} = 1$. 

The protocol involves a sequence of noncommutative operations generated by $\hat{H}_g=\hat{X}^2$\nx{~\cite{Filip2024,box2024}}, $\hat{H}_\lambda=\hat{P}$, implemented via the unitary evolution
$\hat{U}_{\overline{s}} = \prod_{i=0}^N \hat{D}_{x_i} \prod_{j=0}^N \hat{S}_{s_j}$.
Here $\hat{D}_{x_i} = e^{-ix_i\hat{P}}$ denote the momentum displacements and $\hat{S}_{s_j} = e^{-is_j \hat{X}^2}$ are single-mode shearing operations. 
The goal is to estimate the average displacement $\overline{x} = \sum_{i=0}^N x_i/N$, with the mean shearing strength $\overline{s} = \sum_{j=0}^N s_j/N$ treated as an auxiliary parameter.

For this sequence, the nilpotency index between operators $\hat{X}^2$ and $\hat{P}$ is $\mathcal{K} = 1$ while the relevant nested commutator is $[\hat{H}_g,\hat{H}_\lambda]_{(\mathcal{K}=1)} = 2i\hat{X}$,  yielding the local generator $ \hat{h}_{\overline{x}} = N\hat{P} - 2N^2\overline{s}\hat{X}$.
The QFI takes the form:
$ 
    F_{\overline{x}} = 4N^2 \text{Var}[\hat{P}]_{|\Psi\rangle} - 4N^3 \overline{s} \text{Cov}[\hat{X},\hat{P}]_{|\Psi\rangle} + 4N^4 \overline{s}^2 \text{Var}[\hat{X}]_{|\Psi\rangle},
$
where $\text{Cov}[\hat{X},\hat{P}]_{|\Psi\rangle} = \langle \hat{X}\hat{P}+\hat{P}\hat{X}\rangle - 2\langle \hat{X}\rangle \langle\hat{P}\rangle$. For $\overline{s} \neq 0$, the $N^{4}$-scaling term dominates, leading via the QCRB to an RMSE scaling of $\Delta \overline{x} \sim N^{-2}$, surpassing the standard Heisenberg scaling. This enhancement originates purely from the noncommutativity between $\hat{X}^2$ and $\hat{P}$.
However, when the relevant nested commutator is a constant (e.g., $\hat{H}_g=\hat{X}$, $\hat{H}_\lambda=\hat{P}$), ICO enables achieving the same super-Heisenberg scaling $N^{-2}$, a finding corroborated by prior works~\cite{Guo2023,Giulio2020} and consistent with our general result above.

(ii) Infinite nilpotency index $\mathcal{K} \to \infty$.---We now turn to a quantum metrology protocol that achieves exponential enhancement in parameter estimation by a noncommutative process through sequential displacement $\hat{H}_\lambda = \hat{P}$ and squeezing operations $\hat{H}_g = (\hat{a}^\dagger)^2 + \hat{a}^2$. 

\begin{figure}[t]
    \centering   \includegraphics[width=0.6\linewidth]{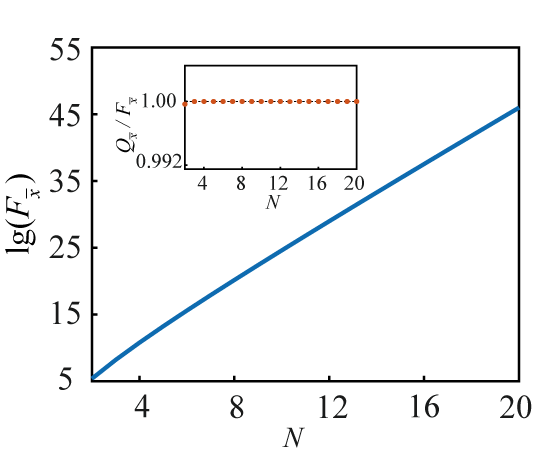}
    \caption{Logarithmic scaling of QFI $F_{\overline{x}}$ with respect to the number of operations $N$ for the noncommutative encoding protocol $\hat{U}_x = \prod_{j=1}^N \hat{D}_{x_j}  \prod_{i=1}^N \hat{S}_{\xi_i}$. The probe state is a coherent state $|\Psi\rangle = \hat{D}(\alpha)|0\rangle$ with $\alpha = 0.3$. Inset: The ratio between the classical Fisher information for a fixed quadrature measurement $Q_{\overline{x}}$ at $\theta=\pi/4$ and the QFI $F_{\overline{x}}$. The ratio consistently remains near unity (dashed line).}
    \label{fig3}
\end{figure}

The parameter of interest is the average displacement $\overline{x} = \sum_{j=1}^N x_j/N$, with an auxiliary squeezing strength $\overline{\xi} = \sum_{i=1}^N \xi_i /N$.
The total unitary evolution is given by $\hat{U}_x = \prod_{j=1}^N \hat{D}_{x_j}  \prod_{i=1}^N \hat{S}_{\xi_i}$, where $\hat{D}_{x_j} = e^{-ix_j \hat{P}}$ and $\hat{S}_{\xi_i} = e^{-i\frac{\xi_i}{2}((\hat{a}^\dagger)^2 + \hat{a}^2)}$. Owing to the noncommutative structure between $\hat{H}_g$ and $\hat{H}_\lambda$, this evolution exhibits an infinite nilpotency index $\mathcal{K} \to \infty$, as evidenced by the unbounded sequence of nested commutators $(\mathrm{ad}_{\hat{H}_g})^n (\hat{H}_\lambda) \neq 0$ for all $n \in \mathbb{N}$. This algebraic structure gives rise to the local generator of the form $\hat{h}_{\overline{x}} =  -N\sinh(N\overline{\xi})\hat{X} + N\cosh(N\overline{\xi})\hat{P}.$
The corresponding QFI reads
\begin{align}
    F_{\overline{x}} & = 4N^2\sinh^2(N\overline{\xi}) \text{Var}[\hat{X}]_{|\Psi\rangle} + 4N^2 \cosh^2(N\overline{\xi}) \text{Var}[\hat{P}]_{|\Psi\rangle} \nonumber \\
    & \quad - 4N^2 \sinh(N\overline{\xi})\cosh(N\overline{\xi}) \text{Cov}[\hat{X},\hat{P}]_{|\Psi\rangle}.
\end{align}

For an initial coherent state $|\alpha\rangle$, the QFI simplifies to $ 2N^2\cosh(2N\overline{\xi})$, demonstrating an exponential scaling $F_{\overline{x}} \simeq N^2e^{2N}$ when $\overline{\xi} > 0$, as shown in Fig.~\ref{fig3}. This exponential enhancement arises from the noncommutative nature of the displacement and squeezing operators. 
Moreover, the final state $|\Psi_{\overline{x}}\rangle = \hat{U}_x|\alpha\rangle$ remains Gaussian, enabling efficient readout via quadrature measurements that nearly saturate the QCRB. The generalized quadrature observable is defined as $\hat{Q} = \hat{X} \cos \theta +\hat{P}\sin\theta$, and its variance in the final state is given by
\begin{equation}
    \Delta^2\hat{Q} = \frac{e^{2N\overline{\xi}}}{4}(1-\sin(2\theta)) + \frac{e^{-2N\overline{\xi}}}{4}(1+\sin(2\theta)).
\end{equation}
The classical Fisher information associated with this measurement reads $Q_{\lambda} = \frac{(\partial_\lambda \langle \hat{Q} \rangle)^2}{\Delta^2 \hat{Q}} + \frac{1}{2} \frac{(\partial_\lambda \Delta^2 \hat{Q})^2}{(\Delta^2 \hat{Q})^2}$. For the sake of brevity, however, we only provide the expressions for the case $\theta = \pi/4$, yielding $Q_{\overline{x}} = N^2 e^{2N\overline{\xi}}$. This demonstrates that a straightforward homodyne measurement protocol can attain precision scaling on par with the QFI.

The encoding processes discussed above can be implemented in a programmable loop-based optical processor capable of executing sequential Gaussian gates. As demonstrated in recent experiments, such platforms support the repeated application of programmable single-mode and multi-step squeezing gates, offering a viable physical system for realizing the proposed noncommutative sensing protocols~\cite{Takeda2021,Takeda2025}. Notably, despite deeper noncommutative structures generally requiring higher circuit depth, more demanding control, and exhibiting stronger vulnerability to decoherence, the infinite-$\mathcal{K}$ limit can be naturally attained in specific platforms.

\textit{Conclusion.---}To summarize, we present a quantum metrology framework harnessing noncommutative operator sequences to surpass standard precision limits. By introducing the nilpotency index $\mathcal{K}$ as a universal characterization to quantify the noncommutativity in the encoding dynamics, we derive scaling laws governing quantum advantage. 
For finite $\mathcal{K}$, the RMSE generically scales as $N^{-(1+\mathcal{K})}$---exceeding the conventional Heisenberg scaling without requiring ICO---provided the $\mathcal{K}$-th order nested commutator remains nonconstant. When this commutator degenerates to a constant, ICO becomes essential to restore optimal scaling. 
When $\mathcal{K} \to \infty$, we show that an exponential enhancement $\Delta \lambda \sim N^{-1} e^{-N}$ can be achieved for a broad class of initial states. 

Moving forward, the quantum-enhanced metrology in our protocol arises from the Lie algebraic relations among the operators. While the present analysis focuses on single-parameter estimation, the proposed scheme can be naturally extended to multi-parameter scenarios. In such settings, the RMSE for composite observables, such as $A:= \overline{\lambda}\overline{g}$, is fundamentally related to the two achievable scaling exponents, especially the worse one~\cite{sm}.
Beyond the sequential encoding protocols, the Lie algebraic structure of noncommutative operators can also be leveraged in critical metrology to surpass quantum speed limits at the criticality~\cite{Cai2021}. This opens an interesting direction for exploring the interplay between noncommutativity and criticality in enhancing sensing performance.

\textit{Acknowledgement.---}This work was supported by Beijing Natural Science Foundation (Grant No. Z240007), National Natural Science Foundation of China (Grants No. 12125402, No. 12534016, No. 12350006, No. 62575232), and Quantum Science and Technology-National Science and Technology Major Project (Grant No. 2024ZD0302401 and No. 2021ZD0301500). Y.X. acknowledges support by the National Cryptologic Science Fund of China (Grant No.~2025NCSF02048).

\textit{Data availability.---}The data that support the findings of this article are openly available~\cite{data}.

\bibliography{ref}

\end{document}